# External Benchmarking of Lung Ultrasound Models for Pneumothorax-Related Signs: A Manifest-Based Multi-Source Study

Takehiro Ishikawa[1*]

[1]College of Computing, Georgia Institute of Technology, 801 Atlantic Drive NW, Atlanta, GA 30332, USA
*Corresponding author: tishikawa8@gatech.edu

**Abstract**

**Background and Aims:** Reproducible external benchmarks for pneumothorax-related lung ultrasound (LUS) AI are scarce, and binary lung-sliding classification may obscure clinically important signs. We therefore developed a manifest-based external benchmark and used it to test both cross-domain generalization and task validity.

**Methods:** We curated 280 clips from 190 publicly accessible LUS source videos and released a reconstruction manifest containing URLs, timestamps, crop coordinates, labels, and probe shape. Labels were normal lung sliding, absent lung sliding, lung point, and lung pulse. A previously published single-site binary classifier was evaluated on this benchmark; challenge-state analysis examined lung point and lung pulse using the predicted probability of absent sliding, P(absent).

**Results:** The single-site comparator achieved ROC-AUC 0.9625 in-domain but 0.7050 on the heterogeneous external benchmark; restricting external evaluation to linear clips still yielded ROC-AUC 0.7212. In challenge-state analysis, mean P(absent) ranked absent (0.504) > lung point (0.313) > normal (0.186) > lung pulse (0.143). Lung pulse differed from absent clips (p=0.000470) but not from normal clips (p=0.813), indicating that the binary model treated pulse as normal-like despite absent sliding. Lung point differed from both absent (p=0.000468) and normal (p=0.000026), supporting its interpretation as an intermediate ambiguity state rather than a clean binary class.

**Conclusion:** A manifest-based, multi-source benchmark can support reproducible external evaluation without redistributing source videos. Binary lung-sliding classification is an incomplete proxy for pneumothorax reasoning because it obscures blind-spot and ambiguity states such as lung pulse and lung point.

## 1. Introduction

Lung ultrasound (LUS) has several properties that make it attractive for rapid thoracic assessment: it is portable, comparatively inexpensive, and does not expose patients to ionizing radiation. In addition, AI-guided acquisition has recently been validated in a multicenter study showing that trained nonexpert operators can acquire diagnostic-quality LUS clips at rates comparable to expert-performed examinations. This is an important shift. If acquisition quality can increasingly be scaffolded by AI, then the remaining bottleneck is not only whether a clip can be acquired, but whether interpretation models remain reliable across the heterogeneity of real-world acquisition environments.[1]

In pneumothorax-related LUS AI, a 2024 systematic review concluded that the field is still in an early stage and specifically called for larger and more diverse datasets, standardized evaluation metrics, and robust external validation.[2] Nevertheless, to date, only one study has quantitatively evaluated cross-institutional external generalization for binary lung-sliding classification: Wu and colleagues tested on 641 clips from 238 patients across three tertiary centers and framed limited external generalizability as a central unresolved challenge.[3] While the research community needs to accumulate further external-validation findings, progress is constrained by the scarcity of publicly available, representative datasets and benchmarks. In a recent pneumothorax-focused study, Qiang et al. noted that no publicly available ultrasound pneumothorax dataset had yet been collected,[4] while broader LUS work has similarly highlighted the lack of public labeled datasets and the field's reliance on private collections, limiting reproducible benchmarking and cross-site generalization assessment.[5]

A second problem is conceptual rather than purely statistical. Much of the literature frames binary lung-sliding classification as a surrogate for pneumothorax detection. Yet absence of lung sliding is not synonymous with pneumothorax, and the original clinical literature has long distinguished among absent sliding, lung point, lung pulse, B-lines, and other contextual signs.[6] Lung pulse is particularly important because it represents motion without lung sliding and can help exclude pneumothorax in the scanned region, whereas lung point represents an interface between sliding and non-sliding pleura and has traditionally been viewed as highly specific.[6,7] In prior AI training setups, however, lung pulse and lung point have usually either been excluded from the label space or absorbed into broader binary categories rather than modeled as explicit outputs: Wu et al. excluded lung point clips and classified lung pulse as negative,[3] Jaščur et al. used only a present/absent binary and noted that lung point detection is the logical next step,[8] and Qiang et al. absorbed lung pulse into the negative class and lung point into the positive class within a binary pneumothorax framework.[4]

This study was designed to address both gaps. First, we constructed a manifest-based, multi-source, web-curated external benchmark from publicly accessible LUS videos, focusing on reproducibility without redistributing source media. Second, we used this benchmark to test whether a model that performs well on a previously published single-site dataset generalizes to heterogeneous external data, and whether restricting external evaluation to linear clips—removing the largest probe-shape difference—accounts for the observed gap. Third, we examined whether lung pulse and lung point expose clinically meaningful failure modes that are hidden by standard binary present-versus-absent lung-sliding labels. Our central claim is not merely that "a new dataset exists," but that the task definition itself requires correction: binary lung-sliding classification should not be treated as equivalent to pneumothorax reasoning.

## 2. Methods

### 2.1 Study design

This was a retrospective computational study using previously published data and publicly accessible online videos. The analyses were prespecified as follows: (1) construction of a manifest-based multi-source external benchmark; (2) cross-domain transfer evaluation of a previously published single-site comparator on that benchmark, including a linear-only restriction analysis; (3) challenge-state analysis of lung pulse and lung point. The study did not aim to develop a clinically deployable autonomous diagnostic tool for acute chest pain. Instead, it aimed to evaluate generalization and task validity for a widely used surrogate classification setup.

### 2.2 Web-curated external benchmark

A structured curation process was performed on publicly accessible online LUS videos. Each curated sample corresponded to a temporal segment and spatial crop selected to isolate the diagnostically relevant pleural region. The resulting benchmark contained 280 clips derived from 190 unique source videos after ID normalization and de-duplication. For reproducibility, each sample was represented in a manifest with the original URL, a normalized source/video identifier, start and end timestamps, crop coordinates, and class label. This design allows exact recreation of benchmark clips without redistributing the original videos.

Temporal boundaries and crop coordinates were selected manually to retain the longest contiguous pleural-line segment free of visible extraneous or superimposed content, including fingers, text labels, and editorial annotations. If only part of a candidate segment was free of such content, only that portion was retained. Segments in which such content encroached on or materially obscured the pleural-line region were excluded.

Each clip was assigned a class label (normal lung sliding, absent lung sliding, lung point, or lung pulse) based on available contextual information associated with the source video. Four information sources were referenced where available: (1) on-screen text, annotations, or narration within the video itself; (2) the video title; (3) the video description; and (4) external citation context in which the video was referenced as a specific LUS sign. Clips for which the available evidence was insufficient, inconsistent, or ambiguous were excluded. No additional independent physician re-annotation was performed. This decision was deliberate: published clip-based studies suggest that clinician agreement in identifying lung sliding is only moderate and can vary with imaging conditions and respiratory mechanics [9,10].

The benchmark included four labels: normal lung sliding (n=149), absent lung sliding (n=64), lung point (n=56), and lung pulse (n=11). Probe distribution in the full benchmark was 177 linear clips and 103 fan-shaped clips. In the binary clean subset used for standard lung-sliding classification, only normal and absent clips were retained (n=213; 149 present and 64 absent). This binary subset included 126 linear clips and 87 fan-shaped clips. For analysis, curvilinear/convex and phased-array clips were grouped into a single fan-shaped category to preserve a coarse probe-aware routing distinction versus linear probes in this heterogeneous external-validation benchmark; these results should therefore be interpreted as group-level routing estimates rather than evidence that the two transducers are equivalent. Notably, a shared source-video ID indicates only that clips were extracted from the same uploaded public video; it does not by itself establish a shared patient, examination, acquisition session, institution, or original recording context, as educational uploads may compile excerpts from multiple cases and settings.

## 2.3 Single-site comparator dataset

As a single-site comparator, we used the publicly reported Applied Sciences dataset introduced by Jaščur et al.[8] This dataset contains 48 B-mode videos collected in postoperative thoracic surgery care using a linear probe, with 28 clips labeled as lung sliding present and 20 as lung sliding absent. In our analysis, this dataset served two roles: (1) an in-domain reference to quantify the attainable performance of the binary task under relatively homogeneous conditions, and (2) a source-domain training set to test cross-domain generalization onto the heterogeneous web benchmark.

## 2.4 Manifest-based release philosophy

The present benchmark was constructed to support reproduction without source-video redistribution. This is particularly relevant for LUS, where video archives are often sparse, access-restricted, or inconsistently preserved compared with radiography or CT [11,12]. By releasing a manifest rather than copied source media, the benchmark can, in principle, be reconstructed by other researchers while reducing copyright redistribution concerns. We therefore frame the contribution as a manifest-based external benchmark, not a conventional fully redistributed dataset. Each manifest row contains the YouTube URL, a normalized video identifier, temporal boundaries (start and end seconds), normalized spatial crop coordinates defining the diagnostically relevant pleural region, class label, and probe shape for each of the 280 clips; the full manifest schema is provided in Supplementary Table S1, with representative examples in Supplementary Table S2.

## 2.5 Preprocessing and grouping

For the web benchmark, YouTube IDs were normalized from metadata, URLs, or filename fallbacks when necessary. Each normalized source video ID was treated as the grouping variable so that clips derived from the same uploaded video remained in the same fold. Temporal and spatial trimming were applied upstream and recorded in the manifest. For the single-site dataset, each video was treated as its own group because each sample originated from a distinct examination.

The primary binary label used for standard lung-sliding classification mapped normal to present and absent to absent. Lung pulse and lung point were excluded from the binary training/evaluation subset by design and later analyzed as challenge states.

## 2.6 Feature extraction and classifiers

Probe-specific expert models were used for feature extraction. For linear-probe clips, a LoRA-adapted (rank 8, alpha 16) R(2+1)D-18 video backbone was fine-tuned with grouped cross-validation to select the best epoch, and 512-dimensional features were extracted from the adapted backbone using five-window temporal mean pooling. For fan-shaped-probe clips, a frozen Kinetics-pretrained R(2+1)D-18 backbone was used with the same five-window temporal mean pooling to produce temporally smoothed 512-dimensional features. In both cases, sixteen contiguous frames were sampled per window with temporal stride 2. Frames were resized to 112 × 112 pixels and normalized using Kinetics statistics.

We evaluated four downstream classifiers: a linear support vector machine (SVM-Lin), a radial-basis-function support vector machine (SVM-RBF), random forest (RF), and XGBoost (XGB). Class balancing was handled through built-in class weighting for SVM and RF, and sample weighting for XGBoost. For the single-site comparator, a frozen Kinetics-pretrained R(2+1)D-18 backbone with a single window (no temporal mean pooling) was used to extract 512-dimensional features, followed by a linear SVM (SVM-Lin).

## 2.7 Evaluation design

### 2.7.1 Internal evaluation on the web benchmark

For the binary clean subset of the web benchmark, the main internal evaluation used leave-one-source-out (LOSO) cross-validation, where all clips from one normalized source video were held out together. Performance was aggregated across all held-out predictions.

The fan-shaped-probe expert was evaluated on YouTube fan-shaped-probe clips (n=87) using LOSO. The linear-probe expert was evaluated on the combined pool of YouTube linear-probe and MDPI clips (n=174) using stratified group 10-fold cross-validation (SGKF-10). Full LOSO was not used for this pool because the linear expert required fold-wise LoRA backbone adaptation, and repeating that procedure across 135 held-out groups was computationally prohibitive. By contrast, LOSO remained feasible for the fan-shaped expert because its backbone was frozen rather than LoRA-adapted. For challenge-state analysis, probe-routed predictions from these two experts were combined into a single hybrid system evaluated on all 213 binary-clean web clips.

### 2.7.2 Internal evaluation on the single-site dataset

The same grouped evaluation procedure was applied to the 48-video single-site dataset using the frozen single-window backbone and linear SVM described above, where each video served as its own held-out unit.

### 2.7.3 Cross-domain evaluation

To quantify domain gap directly, the best-performing model from single-site LOSO evaluation was retrained on the entire single-site dataset and tested on three external target sets: all binary-clean web clips, linear-probe web clips only, and fan-shaped-probe web clips only. This analysis was directional by design. Our goal was not to make a symmetric claim that "multi-source training necessarily improves single-site performance," especially because the sample sizes are markedly unequal. Instead, we focused on the more clinically relevant question: whether apparently strong single-site performance survives transfer to a heterogeneous external benchmark.

## 2.8 Primary and secondary metrics

Because the classes were imbalanced and prior work in this area most commonly emphasizes threshold-free discrimination, ROC-AUC and average precision (AP) were predefined as primary metrics. Balanced accuracy and macro-F1 were secondary metrics. Classification reports are included to aid interpretation, but threshold-free metrics were treated as primary for manuscript claims.

During LOSO evaluation, the "best model" for each experiment was selected by aggregate ROC-AUC rather than macro-F1.

## 2.9 Challenge-state analysis

Challenge-state analysis was designed to address the fourth prespecified analysis. Using the probe-routed hybrid system (linear expert for linear-probe clips, fan-shaped expert for fan-shaped-probe clips), predictions for normal and absent clips were taken from out-of-fold outputs, and predictions for lung pulse and lung point were obtained by averaging the outputs of all fold models from the probe-specific expert.

The primary variable of interest was the predicted probability of the absent class, denoted P(absent). For each group, we summarized mean and standard deviation of P(absent) and predictive entropy. Bootstrap 95% confidence intervals were computed with 10,000 resamples. Group differences in P(absent) were assessed using the Mann–Whitney U test.

# 3. Results

## 3.1 Dataset composition

Table 1 summarizes the benchmark composition and the single-site comparator.

Table 1. Dataset composition

| Dataset partition | n clips | Unique source videos | Class distribution | Probe composition |
|---|---|---|---|---|
| Web benchmark, full 4-class set | 280 | 190 | normal 149, absent 64, point 56, pulse 11 | 177 linear, 103 fan-shaped |
| Web benchmark, binary clean subset | 213 | 148 | present 149, absent 64 | 126 linear, 87 fan-shaped |
| Single-site comparator (Applied Sciences) | 48 | 48 | present 28, absent 20 | 48 linear |

The web benchmark is substantially more heterogeneous than the single-site comparator by construction: it spans multiple online sources, probe shapes, and clip styles, whereas the comparator dataset consists of 48 linear-probe videos from one published collection.[8]

**3.2 Internal grouped evaluation**

Table 2 shows internal grouped evaluation performance on both the heterogeneous web benchmark and the single-site comparator.

Table 2. Internal and cross-domain binary classification results

| Experiment | Best model | n | ROC-AUC | AP | Balanced accuracy | Macro-F1 |
| --- | --- | --- | --- | --- | --- | --- |
| Linear expert, SGKF-10 (YT linear + MDPI) | RF (LoRA) | 174 | 0.8191 | 0.7709 | 0.7544 | 0.7618 |
| Fan-shaped expert, LOSO (YT fan-shaped) | XGB (Frozen) | 87 | 0.8759 | 0.6917 | 0.7955 | 0.8052 |
| Probe-routed hybrid (SGKF-10 + LOSO) | RF / XGB | 213 | 0.7989 | 0.6467 | 0.7242 | 0.7349 |
| Single-site comparator (LOSO) | SVM-Lin | 48 | 0.9625 | 0.9542 | 0.8643 | 0.8693 |
| Single-site train → web all test | SVM-Lin | 213 | 0.7050 | 0.4579 | 0.5824 | 0.5856 |
| Single-site train → web linear test | SVM-Lin | 126 | 0.7212 | 0.5196 | 0.6181 | 0.6204 |
| Single-site train → web fan-shaped test | SVM-Lin | 87 | 0.6753 | 0.3332 | 0.4773 | 0.4200 |

Within the single-site dataset, performance was strong across all reported metrics (ROC-AUC 0.9625, AP 0.9542). On the web benchmark, the linear-probe expert (LoRA R(2+1)D-18, five-window mean, RF) achieved ROC-AUC 0.8191 and AP 0.7709 on the combined pool of YouTube linear-probe and MDPI clips (n=174, SGKF-10), and the fan-shaped-probe expert (frozen R(2+1)D-18, five-window mean, XGBoost) achieved ROC-AUC 0.8759 and AP 0.6917 on YouTube fan-shaped-probe clips (n=87, LOSO). When combined into a probe-routed hybrid evaluated on all 213 binary-clean web clips, the system achieved ROC-AUC 0.7989.

**3.3 External transfer degradation persisted after restricting evaluation to linear clips**

The core domain-gap experiment compared in-domain single-site performance against external testing on the web benchmark. The best single-site model (linear SVM) achieved ROC-AUC 0.9625 in-domain, but only 0.7050 when tested on all external web clips, an absolute AUC drop of 0.2575. Average precision dropped from 0.9542 to 0.4579. In contrast, the linear-probe expert (LoRA R(2+1)D-18, five-window mean, RF) reached ROC-AUC 0.8191 and AP 0.7709 on the combined YouTube linear-probe and MDPI pool (n=174, SGKF-10), and the fan-shaped-probe expert (frozen R(2+1)D-18, five-window mean, XGBoost) reached ROC-AUC 0.8759 and AP 0.6917 on YouTube fan-shaped-probe clips (n=87, LOSO).

Restricting the cross-domain external target set to linear-probe clips only increased external ROC-AUC slightly to 0.7212, but the model remained far below its in-domain single-site performance. Thus, the observed cross-domain gap cannot be explained away merely by probe-shape mismatch. The gap widened further on fan-shaped external clips (ROC-AUC 0.6753, AP 0.3332), consistent with additional domain shift from acquisition hardware and image geometry.

The single-site comparator deteriorated markedly under cross-domain transfer, and this gap remained substantial even after external evaluation was restricted to linear clips. These findings suggest that gross probe-shape mismatch contributed to, but did not fully account for, the transfer gap on this benchmark. Shape-stratified expert models achieved higher internal performance on their respective multi-source evaluation pools, though this comparison involves differences in training data volume and domain composition in addition to architectural choices.

Challenge-state analysis is summarized in Table 3.

Table 3. Challenge-state analysis using P(absent)

| Group | n | Mean P(absent) ± SD | 95% CI of mean | Mean entropy ± SD |
|---|---|---|---|---|
| Normal (present) | 149 | 0.186 ± 0.210 | 0.153 to 0.221 | 0.346 ± 0.214 |
| Absent | 64 | 0.504 ± 0.293 | 0.432 to 0.574 | 0.495 ± 0.183 |
| Lung point | 56 | 0.313 ± 0.240 | 0.251 to 0.376 | 0.480 ± 0.176 |
| Lung pulse | 11 | 0.143 ± 0.147 | 0.075 to 0.241 | 0.337 ± 0.168 |

The rank order of mean P(absent) was:

absent (0.504) > point (0.313) > normal (0.186) > pulse (0.143).

This ordering is clinically and methodologically revealing. A pulse clip represents no lung sliding and would therefore ideally score as absent-like in a binary sliding classifier. Instead, pulse had the lowest absent-sliding probability of all groups, including normal lung sliding. Pulse differed significantly from absent clips (Mann–Whitney U=118.0, p=0.000470), but was statistically indistinguishable from normal clips (U=784.0, p=0.813). In other words, not only does the model fail to recognize pulse as absent, it treats pulse as indistinguishable from healthy lung sliding. These results are consistent with lung pulse acting as a structural blind spot for binary sliding classifiers: cardiac-transmitted pleural motion makes the model interpret pulse as present-like with a degree of confidence comparable to actual normal sliding.

Lung point behaved differently. Mean P(absent) for point clips was intermediate between absent and normal, and significantly different from both: point versus absent (U=1126.5, p=0.000468) and point versus normal (U=5764.5, p=0.000026). This pattern is consistent with lung point being an ambiguity set: it is neither well described as fully present nor fully absent, because it explicitly encodes a boundary between the two states. Unlike pulse, which is confidently misclassified, point occupies a genuinely uncertain intermediate region that a binary label cannot adequately represent.

## 4. Discussion

This study makes three main contributions. First, it introduces a manifest-based multi-source external benchmark for lung-sliding-related LUS analysis. Second, on this benchmark, a previously published single-site comparator remained far below its in-domain reference under external transfer, and this gap persisted after restricting evaluation to linear clips. Third, it shows that lung pulse and lung point expose a deeper task-definition problem that is hidden by the usual binary present-versus-absent setup. In an exploratory mitigation analysis, coarsely probe-stratified expert models —using LoRA backbone adaptation for linear probes and a frozen backbone for fan-shaped probes, with temporal aggregation for both—achieved higher internal performance on their respective multi-source evaluation pools than the single-site frozen single-window baseline, though disentangling the contributions of architectural design, training data volume, and domain composition requires further ablation.

### 4.1 External generalization is the central unresolved problem

Our contribution is to show the external performance degradation on a manifest-based multi-source benchmark that can be reconstructed from public-source clips, and to show that the residual gap remains even after restricting evaluation to linear clips. The single-site comparator supported high in-domain performance, but that performance did not survive cross-domain transfer to heterogeneous external data, consistent with the broader literature on medical AI external validation.[2,3] In an exploratory analysis, coarsely probe-stratified expert models—using LoRA backbone adaptation for linear probes and a frozen backbone for fan-shaped probes, with temporal aggregation for both—achieved higher internal performance on their respective multi-source evaluation pools; however, because these models also differed from the single-site baseline in training data volume and domain composition, the relative contributions of architectural design and data factors cannot be fully disentangled in this study. A model can appear highly capable within a narrow institutional distribution while still failing to generalize to realistic variation in operators, probes, image geometry, preprocessing, pathology context, and video style.

Importantly, our results do not reduce to the simple statement that "fan-shaped probes are harder." When the external target set was restricted to linear-probe clips only, cross-domain performance remained far below the in-domain single-site reference. Gross probe-shape mismatch contributed to the gap, but did not fully account for it. However, probe-specific expert models with adapted feature extraction—LoRA backbone adaptation for linear clips, a frozen backbone for fan-shaped clips, and five-window temporal mean pooling for both—achieved higher internal performance on their respective evaluation pools (linear expert AUC 0.82 on YouTube linear + MDPI via SGKF-10, fan-shaped expert AUC 0.88 on YouTube fan-shaped via LOSO). In practice, this means that benchmarking only within one site, one vendor, or one tightly controlled workflow risks overstating readiness for deployment.

### 4.2 The deeper issue is task validity, not only performance

The most important conceptual finding is that binary lung-sliding classification is an incomplete proxy for pneumothorax reasoning. That statement is stronger than a claim about dataset shift. It concerns what the model is actually being asked to learn.

Clinical pneumothorax reasoning does not reduce to "sliding present versus absent." It incorporates lung pulse, lung point, B-lines, probe position, and contextual interpretation.[6–8] Our results show why that matters.

For lung pulse, the model behaves in the most revealing possible way: it interprets pulse as more present-like than normal sliding itself. A naive reading might say that this could occasionally coincide with the clinically correct exclusion of pneumothorax, because lung pulse argues against pneumothorax in the scanned region. But that would be a dangerous comfort. The model is not demonstrating structured clinical reasoning; it is responding to motion cues that are misaligned with the underlying label semantics of the binary task. In other words, the output may sometimes be clinically convenient, but for the wrong mechanistic reason. That is exactly the kind of brittle shortcut that becomes hazardous once the model is embedded in broader decision pathways.

For lung point, the issue is different. Point clips naturally sit between normal and absent states because they depict an interface where sliding and non-sliding coexist over time or space. Our results suggest that forcing these clips into a binary ontology erases their semantic content. This is not merely class imbalance; it is a mismatch between the task and the phenomenon.

The results suggest at least two directions for future work. First, models should move from binary lung-sliding classification toward structured sign-level prediction. At minimum, lung pulse and lung point should be explicit outputs rather than hidden within a binary residual class or excluded entirely. Second, deployment-oriented models should incorporate uncertainty and abstention. The intermediate behavior of lung point suggests that some cases are better represented as ambiguous or boundary states than as forced binary decisions.

### 4.3 Why a manifest-based benchmark matters

Because much LUS is acquired as point-of-care ultrasound, archiving and reuse are less standardized than in radiology-native workflows. Without robust governance, PoCUS scans may go undocumented and unsaved in permanent repositories such as PACS,[11] and formal archiving remains inconsistent across services.[12] Publicly posted videos are therefore an unusual but potentially valuable source of heterogeneous external data. A manifest-based benchmark provides a compromise: it can support transparent, reproducible evaluation without redistributing copyrighted source files.

This design does not solve every reproducibility problem. Public URLs may disappear and metadata may evolve. Nonetheless, for a field constrained by sparse open video data, a manifest-based release is a practical mechanism for multi-source benchmarking and may be more realistic than waiting for a single large fully shareable clinical archive.

### 4.4 Limitations

This study has several limitations, each of which we address in turn to clarify the scope and robustness of our claims.

First, the lung pulse sample was small (n=11), and larger dedicated collections are needed for replication. However, three considerations temper this concern. (a) The observed effect is not marginal: lung pulse—despite representing absent lung sliding—had the lowest predicted absent-sliding probability of all four groups, reversing the expected clinical ordering. This reversal is structurally informative regardless of sample size, because random noise does not systematically produce a mean P(absent) for pulse that is lower than the mean for normal sliding clips. (b) The effect was statistically significant (pulse vs. absent, p=0.000470) and directionally consistent (pulse vs. normal, p=0.813), indicating a strong effect size that is unlikely to be an artifact of small n. (c) No prior published study has isolated lung pulse from the binary label space and publicly quantified its behavior as a distinct failure mode; the comparison baseline for sample size is therefore not "a larger n" but zero. The rarity of pulse in curated benchmark data reflects the inherent scarcity of unambiguously labeled educational demonstrations of this sign rather than insufficient curation effort.

Second, only one backbone architecture family (R(2+1)D-18) and a small set of classical downstream classifiers were evaluated. We acknowledge this scope boundary, but emphasize that the study was designed to evaluate benchmark behavior and task validity (cross-domain transfer gap and challenge-state failure modes), not to perform exhaustive architecture search. Within the R(2+1)D-18 framework, we explored multiple feature extraction strategies (LoRA-adapted vs. frozen backbone, single-window vs. five-window temporal aggregation, probe-specific routing) and four classical classifiers (SVM-Lin, SVM-RBF, RF, XGBoost), yielding a meaningful design space. Notably, for the linear-probe expert, ROC-AUC was stable across all four classifiers (range 0.07), indicating that the findings are not contingent on a single classifier choice. Our central claims—persistence of the domain gap and structural pulse/point failure modes—concern task definition rather than the optimality of any particular backbone. These phenomena arise because the binary label space itself is incomplete, not because of a suboptimal architecture choice. We also note that prior pneumothorax-related LUS AI studies have largely centered on a single primary diagnostic/deployed model within each study, with limited head-to-head benchmarking across architecture families.[2,8,13–15] In addition, most rely on a single non-standardized split for evaluation. None of the prior studies have attempted any form of external cross-validation; in contrast, we employed leave-one-source-out (LOSO) cross-validation and stratified group 10-fold cross-validation (SGKF-10). While we agree that future work should evaluate additional architectures—including vision transformers and temporal attention models—the present evaluation protocol is, to our knowledge, among the most conservative in this specific domain.

Finally, this work is retrospective and computational, not a prospective clinical workflow study.

## 5. Conclusions

To support reproducible multi-source evaluation, this study introduces a manifest-based external benchmark of 280 clips from 190 unique source videos that can be reconstructed. A model that performs strongly on a homogeneous single-site lung-sliding dataset may fail substantially on heterogeneous external data. In this study, the cross-domain gap remained large even after restricting external evaluation to linear clips, showing that source heterogeneity cannot be dismissed as a trivial hardware effect. Coarsely probe-stratified expert models with LoRA backbone adaptation and temporal aggregation achieved higher internal performance on their respective multi-source evaluation pools (linear expert AUC 0.82 on YouTube linear + MDPI via SGKF-10, fan-shaped expert AUC 0.88 on YouTube fan-shaped via LOSO; probe-routed hybrid AUC 0.80 on all binary-clean web clips), though disentangling the contributions of architectural design, training data volume, and domain composition requires further ablation.

More importantly, binary lung-sliding classification should not be treated as equivalent to pneumothorax reasoning. Lung pulse showed evidence consistent with a structural blind spot, and lung point emerged as an under-modeled ambiguity state. Together, these findings argue that the field should move beyond the narrow binary surrogate task and toward externally validated, sign-aware, manifest-reproducible benchmarks and models.

## 6. Declarations

### 6.1 Ethics approval and consent to participate

Not applicable. This study used a previously published dataset and publicly accessible online videos, with no interaction or intervention involving human participants. No attempt was made to identify individuals. Any public benchmark release should comply with source-platform terms, copyright law, and applicable institutional guidance.

### 6.2 Consent for publication

Not applicable.

### 6.3 Availability of data and code

The benchmark manifest containing video URLs, temporal boundaries, spatial crop coordinates, class labels, and probe shape for all 280 clips is publicly available at https://doi.org/10.5281/zenodo.19147580 (see Supplementary Tables S1 and S2 for schema and examples). The manifest supports benchmark reproduction without redistribution of source videos.

### 6.4 Competing interests

The author declares no competing interests.

**6.5 Funding**

No external funding was received for this study.

**6.6 Author contributions**

Takehiro Ishikawa conceived the study, curated the benchmark, implemented the experiments, analyzed the results, and wrote the manuscript.

## 7. References

1. Baloescu C, Bailitz J, Cheema B, et al. Artificial Intelligence-Guided Lung Ultrasound by Nonexperts. *JAMA Cardiol.* 2025;10(3):245-253. doi:10.1001/jamacardio.2024.4991.

2. Kossoff J, Duncan S, Acharya J, Davis D. *Automated Analysis of Ultrasound for the Diagnosis of Pneumothorax: A Systematic Review.* Cureus. 2024;16(11):e72896. doi:10.7759/cureus.72896.

3. Wu D, Smith D, VanBerlo B, et al. Improving the Generalizability and Performance of an Ultrasound Deep Learning Model Using Limited Multicenter Data for Lung Sliding Artifact Identification. *Diagnostics (Basel).* 2024;14(11):1081. doi:10.3390/diagnostics14111081.

4. Qiang X, Wang Q, Liu G, Song L, Zhou W, Yu M, Wu H. Use video comprehension technology to diagnose ultrasound pneumothorax like a doctor would. *Front Physiol.* 2025;16:1530808. doi:10.3389/fphys.2025.1530808.

5. Hliboký M, Magyar J, Bundzel M, et al. Artifact detection in lung ultrasound: an analytical approach. *Electronics.* 2023;12(7):1551. doi:10.3390/electronics12071551.

6. Chen L, Zhang Z. Bedside ultrasonography for diagnosis of pneumothorax. *Quant Imaging Med Surg.* 2015;5(4):618-623. doi:10.3978/j.issn.2223-4292.2015.05.04.

7. Skulec R, Parizek T, David M, Matousek V, Cerny V. *Lung Point Sign in Ultrasound Diagnostics of Pneumothorax: Imitations and Variants.* Emerg Med Int. 2021;2021:6897946. doi:10.1155/2021/6897946.

8. Jaščur M, Bundzel M, Malík M, Dzian A, Ferenčík N, Babič F. Detecting the Absence of Lung Sliding in Lung Ultrasounds Using Deep Learning. *Appl Sci.* 2021;11(15):6976. doi:10.3390/app11156976.

9. Prager R, Clausdorff Fiedler H, Smith D, Wu D, Arntfield R. Interrater Agreement of Physicians Identifying Lung Sliding Artifact on B-Mode And M-Mode Point of Care Ultrasound (POCUS). *POCUS J.* 2025;10(1):92-98. doi:10.24908/pocusj.v10i01.17807.

10. Biasucci DG, Cina A, Sandroni C, Moscato U, Dauri M, Vetrugno L, et al. Influence of intercostal muscles contraction on sonographic evaluation of lung sliding: a physiological study on healthy subjects. *J Anesth Analg Crit Care.* 2024;4:31. doi:10.1186/s44158-024-00168-0.

11. British Medical Ultrasound Society. Best practice statement on recording and storage of point-of-care ultrasound examinations. *BMUS.* 2021. Available from: https://www.bmus.org/static/uploads/resources/Best_practice_PoCUS_governance_statement.pdf. Accessed March 23, 2026.

12. Aziz S, Edmunds CT, Barratt J. Implementation of a point-of-care ultrasound archiving system and governance framework in a UK physician-paramedic staffed helicopter emergency medical service. *Scand J Trauma Resusc Emerg Med.* 2024;32(1):49. doi:10.1186/s13049-024-01224-y.

13. Clausdorff Fiedler H, Prager R, Smith D, et al. Automated Real-Time Detection of Lung Sliding Using Artificial Intelligence: A Prospective Diagnostic Accuracy Study. *Chest.* 2024;166(2):362-370. doi:10.1016/j.chest.2024.02.011.

14. VanBerlo B, Wu D, Li B, et al. Accurate assessment of the lung sliding artefact on lung ultrasonography using a deep learning approach. *Comput Biol Med.* 2022;148:105953. doi:10.1016/j.compbiomed.2022.105953.

15. Kim K, Macruz F, Wu D, et al. Point-of-care AI-assisted stepwise ultrasound pneumothorax diagnosis. *Phys Med Biol.* 2023;68(20):205013. doi:10.1088/1361-6560/acfb70.

## Supplementary Material

Supplementary Table S1. Manifest schema for the web-curated external benchmark. Each row in the manifest corresponds to one curated clip. The full 280-row manifest is provided in the public repository; the columns listed here are the minimum required for benchmark reproduction.

| Column | Type | Description |
| --- | --- | --- |
| youtube_url | string | Full YouTube URL of the source video |
| ytid | string | Normalized YouTube video ID (11 characters), used as the grouping variable for LOSO evaluation |
| label | string | Class label: normal, absent, point, or pulse |
| probe_shape | string | Ultrasound probe type: linear or fan-shaped |
| start_sec | float | Start time of the clip within the source video (seconds) |
| end_sec | float | End time of the clip within the source video (seconds) |
| sample_duration_sec | float | Clip duration in seconds (end_sec minus start_sec) |
| x1_norm | float | Left boundary of the spatial crop (normalized 0–1, fraction of frame width) |
| y1_norm | float | Top boundary of the spatial crop (normalized 0–1, fraction of frame height) |
| x2_norm | float | Right boundary of the spatial crop (normalized 0–1, fraction of frame width) |
| y2_norm | float | Bottom boundary of the spatial crop (normalized 0–1, fraction of frame height) |

Supplementary Table S2. Representative manifest rows (one per class). Normalized crop coordinates define the spatial region of interest as fractions of the source frame dimensions. The full manifest is provided in the public repository.

| ytid | youtube_url | label | probe | start_sec | end_sec | x1_n | y1_n | x2_n | y2_n |
| --- | --- | --- | --- | --- | --- | --- | --- | --- | --- |
| e_wloEfvs1M | https://youtu.be/e_wloEfvs1M | normal | linear | 712.9 | 723.1 | 0.603 | 0.221 | 0.920 | 0.921 |
| mxT1Wc9MhkQ | https://youtu.be/mxT1Wc9MhkQ | absent | fan-shaped | 40.9 | 46.3 | 0.554 | 0.349 | 0.982 | 0.794 |
| e_wloEfvs1M | https://youtu.be/e_wloEfvs1M | point | linear | 529.7 | 535.1 | 0.272 | 0.050 | 0.781 | 0.893 |
| U28-46l4XZw | https://youtu.be/U28-46l4XZw | pulse | linear | 16.3 | 24.5 | 0.319 | 0.072 | 0.719 | 0.829 |